\RequirePackage{ifpdf}
\ifpdf 
\documentclass[pdftex]{sigma}
\else
\documentclass{sigma}
\fi

\begin{document}

\allowdisplaybreaks

\renewcommand{\PaperNumber}{011}

\FirstPageHeading

\ShortArticleName{Entanglement of  Grassmannian  Coherent States for  Multi-Partite $n$-Level Systems}

\ArticleName{Entanglement of  Grassmannian  Coherent States\\ for  Multi-Partite $\boldsymbol{n}$-Level Systems}

\Author{Ghader NAJARBASHI and Yusef MALEKI}

\AuthorNameForHeading{G.~Najarbashi and Y.~Maleki}

\Address{Department of Physics, University of Mohaghegh Ardabili, Ardabil, 179, Iran}
\Email{\href{mailto:najarbashi@uma.ac.ir}{najarbashi@uma.ac.ir}, \href{mailto:ymaleki@uma.ac.ir}{ymaleki@uma.ac.ir}}

\ArticleDates{Received September 05, 2010, in f\/inal form January 19, 2011;  Published online January 24, 2011}

\Abstract{In this paper, we investigate the entanglement of multi-partite
Grassmannian coherent states (GCSs)  described by  Grassmann
numbers for $n>2$  degree of nilpotency. Choosing an appropriate weight function, we show that it is
possible to construct some well-known entangled pure states,
consisting of {\bf GHZ}, {\bf W}, Bell, cluster type and
bi-separable  states, which are obtained by integrating over tensor
product of GCSs. It is shown that for three level systems, the
Grassmann creation and
 annihilation operators
$b$ and  $b^\dag$ together with $b_{z}$ form a closed deformed
algebra, i.e.,
 $SU_{q}(2)$  with $q=e^{\frac{2\pi i}{3}}$, which is useful to construct
entangled qutrit-states. The same argument  holds for three level
squeezed states. Moreover combining the
Grassmann and bosonic coherent states we
 construct maximal entangled super coherent states.}

\Keywords{entanglement; Grassmannian variables; coherent states}

\Classification{81R30; 15A75; 81P40}

\section{Introduction}

Quantum entanglement has been recognized as the vital resource for
the applications of quantum information and quantum computation. The
emergence of entanglement is a fundamental dif\/ference between
classical and quantum composite systems. Consequently The question
of entanglement of composite systems has been intensively studied in
last years \cite{Nielsen1,petz1}.

In the same direction, the growth of research in theoretical physics
and quantum optics has revealed the importance of the coherent
states and hence the entanglement of the bosonic, su(2) and su(1,1)
coherent states has been widely investigated
\cite{Enk1,Enk2,Fujii1,Naj4,Wang1,Wang2,Wang3,Wang4,Wang5}. On the
other hand, studying the entanglement properties of Grassmannian
coherent states (GCSs) which is def\/ined as a eigenstate of the
annihilation operator
 with Grassmannian eigenvalue \cite{Majid,Cabra,Naj1}, remains as a challenging problem of quantum information theory \cite{Duff, Khanna, Castellani, Naj2}.
In~\cite{Naj3} we have investigated the relation between entanglement and
fermionic coherent states~\cite{Glauber}.

Aim of the present work is to generalize our previous attempt in~\cite{Naj3} to multi-level GCS. Choosing some appropriate weight
functions, we show that it is possible to construct some entangled
pure states, consisting of {\bf GHZ}, {\bf W}, Bell, cluster type
and bi-separable  states, by tensor product of GCSs. It is shown
that for three level systems, the Grassmann creation and
annihilation operators $b$ and  $b^\dag$ together with $b_{z}$ form
a closed deformed algebra, i.e., $SU_{q}(2)$ with $q=e^{\frac{2\pi
i}{3}}$.  Based on this algebra and corresponding GCS  we construct
some entangled qutrit-states. Similar discussion is made for three
level Grassmannian squeezed states.  Moreover combining the
Grassmannian and bosonic coherent states we construct maximal
entangled super coherent state.

The paper is organized as follows. In Section~\ref{section2}, the GCS for $n$
level system is introduced. In Section~\ref{section3}, the entangled GCS
 is studied and explicit examples of multi-qubit entangled states such as Bell,
 $\mathbf{W}$, $\mathbf{GHZ}$, bi-separable and cluster type states are
constructed. Moreover, entang\-led qutrit states based on coherent and
squeezed states are presented and then the discussion is generalized
for entangled multi-qudit states. Finally combining Grassmannian and
bosonic coherent states we construct maximal entangled super
coherent state. A brief conclusion is given in Section~\ref{section4}.

\section{Grassmannian coherent states}\label{section2}

 Grassmann variables and their applications have been discussed
in~\cite{Majid,Glauber,Kerner,Filip,Isa,Cug,ILInski}. Here we
review the properties which are particularly appealing for our
purposes. Grassmann algebra is generated by  variables like
$\theta_{i}s$ which have the following properties
\begin{gather*}
    \theta_{i}\theta_{j}=q \theta_{j}\theta_{i}, \quad
i,j=1,2,\dots,\quad  i<j, \qquad
 \theta_{i}^n=0 , \qquad q=e^{\frac{2\pi i}{n}}. 
\end{gather*}
Likewise, for the Hermitian conjugate of $\theta$,
$\theta^\dag=\bar{\theta}$, we have
\begin{gather*}
    \bar{\theta}_{i}\bar{\theta}_{j}=q  \bar{\theta}_{j}\bar{\theta}_{i} , \quad
 i<j, \qquad
 \bar{\theta}_{i}^n=0 .
  \end{gather*}
One has the Berezin's rule of integration as follows
\begin{gather*}
    \int d\theta \theta^{k}= \int d\bar{\theta}
\bar{\theta}^{k}=\delta_{k,n-1},
\end{gather*}
where $k$ is a positive integer. Moreover we have the relations
\begin{alignat*}{3}
&  \theta d \bar{\theta} = q  d \bar{\theta} \theta , \qquad & & \bar{\theta} d \theta = q   d \theta \bar{\theta}, &  \\
&  \theta d \theta = \bar{q}  d \theta \theta , \qquad & & \bar{\theta} d \bar{\theta} = \bar{q}  d \bar{\theta} \bar{\theta}, &  \\
&  d \theta d \bar{\theta} = \bar{q} d \bar{\theta} d\theta , \qquad&&  \theta  \bar{\theta} = \bar{q}   \bar{\theta} \theta  . &
\end{alignat*}
The quantization relations between number states $\{{|m\rangle}, \
m=0,1,2,\dots,n-1\}$ and Grassmann variables $\theta$, $\bar{\theta}$ are
\begin{gather*}
 \theta   |m\rangle = q^{^{m-1}}   |m\rangle\theta ,\qquad
  \theta   \langle m|  = \bar{q}^{^{m-1}}
\langle m|  \theta .
\end{gather*}
By def\/inition, GCS denoted by $|\theta\rangle_{n}$, is the
eigenstate of annihilation operator $b$ with eigenva\-lue~$\theta$ that is
\begin{gather*}
b|\theta\rangle_{n} =\theta   |\theta\rangle_{n},
\end{gather*}
where $\theta$ is a complex Grassmann variable  with the following
relations
\begin{gather*}
  {[\theta , b^\dag]}_{q}={[b, \theta }]_{q}=0,
\end{gather*}
where ${[A,B]}_{q}:=AB-qBA$.
One may tempt to def\/ine  the
annihilation operator $b$ as
\begin{gather*}
b=\sum_{m=0}^{n-1}\sqrt{m+1}  |m\rangle\langle m+1|.
\end{gather*}
Hence $|\theta\rangle_{n}$ can be def\/ine as below
\begin{gather}\label{coherentform1}
|\theta\rangle_{n}=\sum_{m=0}^{n-1}
\frac{\bar{q}^{^{\frac{m(m+1)}{2}}}}{\sqrt{m!}}  \theta ^{m}
|m\rangle=: e^ {( b^\dag \theta)}
|{0}\rangle.
\end{gather}
One may regard  $D(\theta):=e^ {( b^\dag \theta)}$ as the
displacement operator for GCS.

\section{Entangled Grassmann coherent states}\label{section3}

Consider a linear combination of tensor product of Grassmannian
coherent states as follows
\begin{gather*}
|\psi\rangle=\sum_{i_1,i_2,\dots,i_n}f_{i_1,i_2,\dots,i_n}|{\theta}_{i_1}\rangle|{\theta}_{i_2}\rangle \cdots |{\theta}_{i_n}\rangle.
\end{gather*}
Some times, if we take a proper form of the tensor product of
Grassmannian coherent states, it is possible to f\/ind an appropriate
weight function to make the right hand side of the following
equation to be a maximally entangled state like  Bell, cluster type,
{\bf{GHZ}} and {\bf{W}} states, i.e.,
\begin{gather}\label{EFCS1}
\int  d{\theta}_{i_1} d{\theta}_{i_2}\cdots
d{\theta}_{i_n}w({\theta}_{i_1},\dots,{\theta}_{i_n})|\psi\rangle
=|\gamma\rangle,
\end{gather}
where $ w({\theta}_{i_1},\dots,{\theta}_{i_n})$  is a proper weight
function, and the state $|\gamma\rangle,$ is maximally entangled
state. We note that the weight functions is not unique and of course
for a given state there may not be such a function at all. Here
there is no need to normalize GCSs
 since it can be included in  the weight functions.

\subsection{Multi-qubit states}
Two level Grassmannian coherent state can be written as
\begin{gather*}
|\theta\rangle=|0\rangle-\theta|1\rangle.
\end{gather*}
At f\/irst, we consider simple cases that yield maximally entangled
Bell states as follows
\begin{gather*}
\int d\theta   \left(\frac{\mp1}{2\sqrt{2}}\right)[|\theta\rangle
|\pm\theta\rangle-|-\theta\rangle
|\mp\theta\rangle]=\frac{1}{\sqrt{2}}(|01\rangle\pm|10\rangle)=
|{\Psi}^{\pm}\rangle,
\\
\int d\bar{\theta} d\theta \left (\frac{\pm1}{\sqrt{2}}
e^{\pm\theta\bar{\theta}}
\right)|\bar{\theta}\rangle|\theta\rangle=\frac{1}{\sqrt{2}}(|00\rangle\pm|11\rangle)
=|{\Phi}^{\pm}\rangle.
\end{gather*}
One gets the general form of the {\bf W} state as follows
\begin{gather*}
\int
d\theta\left(\frac{-1}{\sqrt{n}}\right)\underbrace{|\theta\rangle|\theta\rangle \cdots |\theta\rangle}_{n\ {\rm times}}=\frac{1}{\sqrt{n}}(|100\dots 0\rangle+|010\dots 0\rangle+
\cdots +|0\dots 001\rangle)  =|{\bf W}^{(n)}\rangle.
\end{gather*}
Likewise, we can construct the general form of {\bf GHZ} state as
follows
\begin{gather*}
\int
d\theta_{1}d\theta_{2}\cdots d\theta_{n}w|\theta_{n}\rangle|\theta_{n-1}\rangle \cdots |\theta_{1}\rangle=\frac{1}{\sqrt{2}}
(|00\dots 0\rangle+|11\dots 1\rangle)=|{\bf GHZ}^{(n)}\rangle,
\end{gather*}
where
\begin{gather*}
w=\frac{1}{\sqrt{2}}\big((-1)^{[\frac n2]}+\theta_{n}\theta_{n-1}\cdots \theta_{1}\big).
\end{gather*}
Note that  unlike ${\bf W}^{(n)}$, we can not construct
${\bf GHZ}^{(n)}$ using just one  Grassmann variable for $n>2$. The
amount of entanglement of these states can be evaluated  using the
purity which is def\/ined as $p(|\psi\rangle)=\frac{2}{n}\sum_i
{\rm tr}\, \rho_i^2-1$, where $\rho_i$ is reduced density matrix of
 qubit~$i$~\cite{Viola}.
For ${\bf GHZ}^{(n)}$ state the purity is zero and  for
${\bf W}^{(n)}$ state reads $(\frac{n-2}{n})^2$, which implies that
${\bf GHZ}^{(n)}$  state
 is  maximally entangled while in the limit $n\rightarrow\infty$
the purity goes to~1 for ${\bf W}^{(n)}$ hence it  become
separable. The other set of important entangled states are cluster
type states~\cite{Munhoz}. Here we consider one example of these
states as follows
\begin{gather*}
|{\rm CLUSTER}^\pm\rangle=\frac{1}{2}(\pm|0000\rangle+|0011\rangle+|1100\rangle\mp|1111\rangle).
\end{gather*}
If we take
$|\psi\rangle=|\theta_{1}\rangle|\theta_{2}\rangle|\theta_{3}\rangle|\theta_{4}\rangle$,
in equation~(\ref{EFCS1}), then we get
\begin{gather*}
\int d{\theta}_{1}\cdots
d{\theta}_{4}w_{({\rm CLUSTER}^\pm)}|\theta_{1}\rangle|\theta_{2}\rangle|\theta_{3}\rangle|\theta_{4}\rangle
=|{\rm CLUSTER}^\pm\rangle,
\end{gather*}
where
\begin{gather*}
w_{({\rm CLUSTER}^\pm)}=\frac{1}{2}(\pm\theta_{4}\theta_{3}\theta_{2}\theta_{1}
+\theta_{2}\theta_{1}+\theta_{4}\theta_{3} \mp1).
\end{gather*}
Taking the same method, one can construct the other cluster type
states  with appropriate weight functions.

\subsection[Multi-qutrit states and $SU_{q}(2)$ deformed algebra]{Multi-qutrit states and $\boldsymbol{SU_{q}(2)}$ deformed algebra}
Multi-qutrit states  are three level systems. Therefore, we take the
following   bases to describe these states
\begin{gather*}
 |0\rangle \equiv
 \left(\begin{array}{c}
    1\\
    0 \\
    0
 \end{array}\right),
\qquad
|1\rangle \equiv
 \left(\begin{array}{c}
    0\\
    1 \\
    0
 \end{array}\right),
\qquad
|2\rangle \equiv
 \left(\begin{array}{c}
    0\\
    0 \\
    1
 \end{array}\right).
\end{gather*}

The associated Grassmann numbers satisfy $\theta^3=0$. So, the
following quantization relations hold
\begin{alignat*}{3}
& \theta   |0\rangle = \bar{q}   |{0}\rangle \theta  ,   \qquad &&  \langle{0}|  \bar{\theta} = q   \bar{\theta}  \langle0|, & \\
&   \theta  |{1}\rangle =   |{1}\rangle  \theta  ,   \qquad &&  \langle{1}|  \bar{\theta} =    \bar{\theta}  \langle{1}|, & \\
&  \theta  |{2}\rangle = q   |{2}\rangle   \theta   ,   \qquad &&  \langle{2}| \bar{\theta} = \bar{q}  \bar{\theta}  \langle{2}|,  &
\end{alignat*}
where $q=e^{\frac{2\pi i}{3}}$. Now, we are going to derive an
algebra under which three level GCSs can be constructed. For this
purpose, consider the explicit form of the three level annihilation
and creation operators $b$, $b^{\dag}$ and $b_{z}$ as follows
\begin{gather*}
b:=  |0\rangle\langle{1}|+\sqrt{{2}}
|{1}\rangle\langle{2}|,
\qquad b^{\dag}:= |{1}\rangle\langle{0}|+\sqrt{{2}}
|{2}\rangle\langle{1}|,
\qquad
b_{z}:=[b,b^{\dag}]_{q}:=bb^{\dag}-q b^{\dag}b.
\end{gather*}
Using the commutation relation between operators $b$ and $b_{z}$, we
get
\begin{gather}\label{bzb}
[b_{z},b]_{q}=\big({1}-{2}q+q^2 \big)|{0}\rangle\langle{1}|+\big({2}-q+2q^
{2}\big)\sqrt{{2}}|{1}\rangle\langle{2}|.
\end{gather}
If  the right hand side of the above equation is taken to be
proportional to $b$,  we must have
\begin{gather*}
\big({1}-{2}q +q^2 \big)=\big({2}-q +{2}q^2\big),
\end{gather*}
or equivalently
\begin{gather}\label{eqvalentely}
(1+q+q^2)=0 \quad \Rightarrow \quad q=e^{\frac{2\pi i}{3}}.
\end{gather}
Thus the equation~(\ref{bzb}) reduces to
\begin{gather*}
{[b_{z},b]}_{q}=-3q b.
\end{gather*}
Developing the same method to the commutator of $b^{\dag}$ and
$b_{z}$ the condition (\ref{eqvalentely}) is obtained again. Hence
we have
\begin{gather*}
[b,b^{\dag}]_{q}=b_{z},\qquad
  {[b_{z},b]}_{q}=-3qb,\qquad
  {[b^{\dag},b_{z}]}_{q}=-3qb^{\dag}.
\end{gather*}
As we see this algebra is closed and it is reminiscent of the
$SU(2)$ algebra, and we call  it deformed $SU_{q}(2)$  algebra. The
coherent state for three level system is
\begin{gather*}
  |\theta\rangle= |{0}\rangle+\bar{q}
\theta |{1}\rangle+ \frac{1}{\sqrt{{2}}}
\theta^2   |{2}\rangle  = \big(1+ b^{\dag} \theta    + \tfrac{1}{2}  \theta^2
  {b^{\dag}}^2 \big)|0\rangle.
\end{gather*}
Using the second equality we rewrite three level GCS in terms of the
exponential function, def\/ined in equation~(\ref{coherentform1}) as
follows
\begin{gather*}
|\theta\rangle= e^ {(b^{\dag} \theta)}  |0\rangle.
\end{gather*}
Now we will use this coherent state to construct some  entangled
qutrit states. To do so, we start with generalized Bell states for
three level system~\cite{Fujii2}. These states are
\begin{gather*}
|\psi_\pm\rangle=\frac{1}{\sqrt{3}}(|00\rangle\pm|11\rangle+|22\rangle),
\qquad
|\varphi_\pm\rangle=\frac{1}{\sqrt{3}}(|02\rangle\pm|11\rangle+|20\rangle).
\end{gather*}
In order to obtain the above states,  we take the state
$|\theta_1\rangle|\theta_2\rangle$. Therefore we have
\begin{gather*}
\int d{\theta}_{1} d{\theta}_{2}
\frac{1}{\sqrt{3}}\big(\theta_2^2\theta_1^2\pm {q}^2
\theta_1\theta_2 +2q\big)
|\theta_{1}\rangle|\theta_{2}\rangle=|\psi_\pm\rangle ,
\end{gather*}
likewise
\begin{gather*}
\int d{\theta}_{1} d{\theta}_{2}
\frac{1}{\sqrt{3}}\big(\sqrt{2}\theta_1^2\pm q^2
\theta_1\theta_2 +\sqrt{2}\theta_2^2\big)
|\theta_{1}\rangle|\theta_{2}\rangle=|\varphi_\pm\rangle.
\end{gather*}
Of course the weight function may be chosen  in a way that it yields
contraction, this means that we may  have MES in the subspaces. The
following examples are of this type
\begin{gather*}
\int d{\theta}_{1} d{\theta}_{2}
\frac{1}{\sqrt{2}}\big(\theta_2^2\theta_1^2 \pm2q\big)
|\theta_{1}\rangle|\theta_{2}\rangle=\frac{1}{\sqrt{2}}(|00\rangle\pm|22\rangle),
\\
\int d{\theta}_{1}
d{\theta}_{2}\frac{1}{\sqrt{2}}\big(\theta_2^2\theta_1^2\pm
q^2 \theta_1\theta_2\big)
|\theta_{1}\rangle|\theta_{2}\rangle=\frac{1}{\sqrt{2}}(|00\rangle\pm|11\rangle),
\end{gather*}
which is comparable with hole burning of an atomic coherent state
prepared for a collection of~$N$ two-level atoms~\cite{Gerry}. One
may construct bi-separable states for three partite  systems,  for
instance
\begin{gather*}
\int d{\theta}_{1} d{\theta}_{2}d{\theta}_{3}
\frac{1}{\sqrt{3}}\big(\theta_3^2\theta_2^2\theta_1^2\pm
\bar{q}\theta_1^2\theta_2\theta_3\big)
|\theta_{1}\rangle|\theta_{2}\rangle
|\theta_{3}\rangle=|0\rangle\otimes\frac{1}{\sqrt{2}}(|00\rangle\pm|11\rangle).
\end{gather*}
By changing the weight functions, one may get some  other entangled
states as well. For example, we can create all MESs introduced in~\cite{Chunfang1}. Here we give an example  as follows
\begin{gather*}
\int d{\theta}_{1} d{\theta}_{2}
\frac{1}{\sqrt{3}}\big(q^2\theta_{1}^2\theta_{2}+\sqrt{2}\omega\theta_{1}+\sqrt{2}\omega^2\theta_{2}^2\big)
|\theta_{1}\rangle|\theta_{2}\rangle=|\Psi^{(2)}_2\rangle,
\end{gather*}
where $ \omega^3=1$ and
\begin{gather*}
|\Psi^{(2)}_2\rangle=\frac{1}{\sqrt{3}}(|01\rangle+\omega|12\rangle+\omega^2|20\rangle).
\end{gather*}

\subsection{Grassmannian qutrit squeezed states}

For three level system, it is possible to def\/ine squeezing operator
associated to Grassmann  number as follows
\begin{gather}\label{sq-operator}
S(\xi)=\exp\left[\frac{1}{2}\big(\xi  b^{\dag^2}- \bar\xi b^{2}\big)\right].
\end{gather}
Here  instead of $\theta$  we use $\xi,$ for three level squeezed
state. Noting that in the three level system,  the operators $b^{3}$
and $b^{\dag^3}$ vanish and hence the expansion of squeezing
operator becomes
\begin{gather*}
S(\xi )=I+\frac{1}{2}\big(\xi  b^{\dag^2}- \bar{\xi }
b^2\big)-\frac{\bar{q}}{8} \xi \bar{\xi}
\big(b^{\dag^2}b^2+qb^2b^{\dag^2}\big).
\end{gather*}
 It is remarkable that the operators $b^{2}$ and $b^{\dag^2}$, used in
squeezing operator for three level system, together with $b'_{z}$
obey a closed algebra as follows
\begin{gather*}
 [b^{\dag^2},b^2]:=b'_{z},\qquad
  {[b'_{z},b^2]}=-8b^2,\qquad
    {[b'_{z},b^{\dag^2}]}=8b^{\dag^2}.
\end{gather*}
Grassmannian squeezed states by def\/inition can be obtained by
applying squeezing opera\-tor~$S(\xi)$ on the vacuum state
$|{0}\rangle$, i.e.
\begin{gather*}
|\xi\rangle=S(\xi )|{0}\rangle.
\end{gather*}
Therefore for three level system we get the squeezed state as
\begin{gather}
|\xi\rangle=|{0}\rangle+\frac{1}{\sqrt{2}} \xi|{2}\rangle
-\frac{1}{4}\xi\bar{\xi}|{0}\rangle
 =\left(1-\frac{1}{4}\xi\bar{\xi}\right)|{0}\rangle+ \frac{1}{\sqrt{2}}
\xi|{2}\rangle. \label{squeez}
\end{gather}
Therefore we can  construct entangled state,
$\frac{1}{\sqrt{2}}(|00\rangle+|22\rangle)$ just by one Grassmannian
squeezed state $|\xi\rangle$ which such construction is not possible
by  GCS with one Grassman number, i.e.,
\begin{gather}\label{entangl-squeezed}
\int d\bar{\xi}d\xi   \frac{1}{\sqrt{2}}\big(2\bar q \bar\xi^2-16\bar
{q}-2\bar q \xi\bar\xi+\bar\xi^2\xi^2\big)|\xi\rangle
|\xi\rangle=\frac{1}{\sqrt{2}}(|00\rangle+|22\rangle).
\end{gather}
Now, we construct entangled state using tensor product of the
coherent  and squeezed states as follows
\begin{gather*}
\int d\theta
(q+\theta)\widetilde{|\theta\rangle}|\theta\rangle=\frac{1}{\sqrt{2}}(|02\rangle+|20\rangle),
\end{gather*}
where $\widetilde{|\theta\rangle}$ is squeezed state~(\ref{squeez}),
(which $\xi$  is replace with $\theta$) and   ${|\theta\rangle}$ is
GCS for  three level system. It is notable that, in this case we
have used one Grassmann number and obtained the state
$\frac{1}{\sqrt{2}}(|02\rangle+|20\rangle)$, which is not possible
to be obtained using tensor product of GCSs  with just one Grassmann
number, i.e., there is no weight function  $w$ such that $\int
d\theta
w|\theta\rangle|\theta\rangle=\frac{1}{\sqrt{2}}(|02\rangle+|20\rangle)$.

\subsection{Multi-qudit entangled states}
We should mention that it is easy to generalize this method to any
$Z_{n}$ graded  GCS. To this aim let us start with the general form
of the GCS
\begin{gather*}
|\theta\rangle_{n}=\sum_{m=0}^{n-1}
\frac{\bar{q}^{^{\frac{m(m+1)}{2}}}}{\sqrt{m!}}  \theta ^{m}
|m\rangle,
\end{gather*}
we product two GCSs with dif\/ferent Grassmann numbers as
\begin{gather*}
|\theta_1\rangle_{n}|\theta_2\rangle_{n}=\sum_{i,j=0}^{n-1}c_{ij}
\theta_1^i\theta_2^j|i\rangle|j\rangle,
\qquad \mbox{where} \qquad
c_{ij}=\frac{q{^{\frac{(j-i)-(i+j)^2}{2}}}}{\sqrt{i!j!}}.
\end{gather*}
 In~\cite{Ichikawa} it has been shown that the symmetric states are either globally entangled
or fully se\-pa\-rable with all the constituent systems having identical
states, whereas antisymmetric states are globally entangled. By
globally entangled we mean that  the state remains entangled across
any bi-partition. On the other hand  a state is fully separable if
it remains separable across all bi-partitions.
 Therefore  we may determine the weight function in a way that the
obtained state becomes symmetric MES after integration, i.e.,
\begin{gather}\label{gen2}
\int d\theta_1d\theta_2w|\theta_1\rangle_{n}|\theta_2\rangle_{n}
=\frac{1}{\sqrt{n}}\sum_{i=0}^{n-1} |i\rangle|i\rangle,
\\
\label{gen1}
\int d\theta_1d\theta_2w\sum_{i,j=0}^{n-1}c_{ij}
\theta_1^i\theta_2^j|i\rangle|j\rangle
=\frac{1}{\sqrt{n}}\sum_{i=0}^{n-1} |i\rangle|i\rangle.
\end{gather}
Now consider  the general form of Grassmannian weight function
\begin{gather*}
w=\sum_{k,l=0}^{n-1}w_{k,l}\theta_1^{k}\theta_2^{l}.
\end{gather*}
Putting this weight in equation~(\ref{gen1}) and taking into account the
quantization  and the integration rules of generalized Grassmannian
variables  we have
\begin{gather}
\sum_{k,l=0}^{n-1}\sum_{i,j=0}^{n-1}c_{ij}w_{k,l}\int
d\theta_1d\theta_2\theta_1^{k}\theta_2^{l}
\theta_1^i\theta_2^j|i\rangle|j\rangle
=\frac{1}{\sqrt{n}}\sum_{i=0}^{n-1} |i\rangle|i\rangle,\nonumber
\\
\sum_{k,l=0}^{n-1}\sum_{i,j=0}^{n-1}c_{ij}w_{k,l}q^{kl+jk+ij}\int
d\theta_1d\theta_2\theta_2^{l+j} \theta_1^{i+k}|i\rangle|j\rangle
=\frac{1}{\sqrt{n}}\sum_{i=0}^{n-1} |i\rangle|i\rangle,\nonumber
\\
\label{qudit1}
\sum_{k,l=0}^{n-1}\sum_{i,j=0}^{n-1}c_{ij}w_{k,l}q^{kl+jk+ij}
\delta^{l+j}_{n-1}\delta^{i+k}_{n-1}|i\rangle|j\rangle
=\frac{1}{\sqrt{n}}\sum_{i=0}^{n-1} |i\rangle|i\rangle,
\end{gather}
where  the symbol  $\delta^i_{j}$ is the usual Kronecker delta. We
note that
\begin{gather}\label{qudit2}
\delta^{l+j}_{n-1}\delta^{i+k}_{n-1}\neq0 \quad \Longrightarrow
\quad {l+j}={n-1}={i+k}.
\end{gather}
The equation (\ref{qudit1}) gives the right hand side MES if the
terms with $i\neq j$ vanish, which due to the equation (\ref{qudit2})
implies  that $w_{k,l}=0$ for $k\neq l$.
 With this explanation the equation~(\ref{qudit1}) reduces to
\begin{gather*}
\sum_{i=0}^{n-1}c_{ii}w_{n-1-i,n-1-i}q^{(n-1-i)(n-1)+i^2}
|i\rangle|i\rangle =\frac{1}{\sqrt{n}}\sum_{i=0}^{n-1}
|i\rangle|i\rangle.
\end{gather*}
Thus{\samepage
\begin{gather*}
w=\frac{1}{\sqrt{n}}\sum_{k=0}^{n-1}c_{(n-1-k),(n-1-k)}^{-1}{\bar{q}}^{^{k(n-1)+(n-1-k)^2}}\theta_1^{n-1-k}\theta_2^{n-1-k}.
\end{gather*}
With this weight the equation~(\ref{gen2}) holds.}

We note that   any attempt for f\/inding f\/inite deformed $SU_{q}(2)$
algebra, as for qutrit case, for more than three level systems will
cease to exist. Actually these generators
 do not form~\emph{finite} closed algebra at all.
 In the previous subsection we def\/ined squeezed state for three level
 system by action the squeezing operator~(\ref{sq-operator}) on vacuum state $|0\rangle$.
 One may tempt to def\/ine the squeezed state for general qudit states. Here is a point. Instead of using squeezing operator~(\ref{sq-operator}), we use $S(\xi)=e^{(\xi b^{\dag^2})}$ to def\/ine squeezed state
\begin{gather*}
|\xi\rangle=e^{(\xi b^{\dag^2})} |{0}\rangle.
\end{gather*}
  This makes  the coef\/f\/icients  change in $|\xi\rangle$. For
 instance if we take $S(\xi)=e^{(\xi b^{\dag^2})}$, then its associated  qutrit squeezed state
 becomes  $|{0}\rangle+\xi|{2}\rangle$,  which  up to the coef\/f\/icients is the equation~(\ref{squeez}). But the construction  of  maximally entangled qutrit states  go to the same state  in (\ref{entangl-squeezed})
 no matter how we take the squeezing operator here, i.e.,
\begin{gather*}
\int d\xi   \frac{1}{\sqrt{2}}\big(1+\xi^2\big)|\xi\rangle
|\xi\rangle=\frac{1}{\sqrt{2}}(|00\rangle+|22\rangle).
\end{gather*}
 As was clearly seen in qutrit case, the squeezed state
 is superposition of even number states $|{0}\rangle$ and $|{2}\rangle$. This is the case in general.
Now we use this operator to develop equation (\ref{sq-operator}) to
qudit case. The general squeezed state is
\begin{gather*}
|\xi\rangle=e^{(\xi b^{\dag^2})} |{0}\rangle=\sum_{i=0}^{n-1}
\frac{\bar{q}^{^{{i(i-1)}}}}{{i!}} \xi ^{i} |2i\rangle.
\end{gather*}
Thus considering the product of two squeezed states
\begin{gather*}
|\xi\rangle|\xi\rangle=\sum_{i,j=0}^{n-1} d_{i,j} \xi ^{i+j}
|2i\rangle|2j\rangle,
\end{gather*}
with
$d_{i,j}=\frac{\bar{q}^{^{{i(i-1)}+{j(j-1)}+(2j-1)j}}}{{i!j!}}$,
one can create the maximally entangled state  as follows
\begin{gather*}
\int d\theta   w
|\xi\rangle|\xi\rangle=\frac{1}{\sqrt{n}}\sum_{k=0}^{n-1}
|2k\rangle|2k\rangle,
\end{gather*}
where the general form of the one-variable weight function  is
\begin{gather*}
w=\sum_{m=0}^{n-1} w_{m} \xi ^{m}.
\end{gather*}
Similar to the $n$ level qudit coherent states we can f\/ind the
weight function as
\begin{gather*}
w=\frac{1}{\sqrt{n}}\sum_{i=0}^{n-1}  \frac{1}{d_{ii}}
\xi^{n-2i-1}.
\end{gather*}

There is   rather tight relation between the  construction of
entangled states of this work and~\cite{Mandilara} where
they used the same method to characterize the entanglement by
polynomials of fermionic nilpotent raising operators  $\sigma^+$,
acting on a reference vacuum state. The dif\/ference is that we use
Grassmann anticommuting variables instead of Clif\/ford nilpotent
variables.  For example in two qubit case one may  take
\begin{gather*}
F(\sigma_1^+,\sigma_2^+)=a_0+a_1\sigma_1^+ +a_2\sigma_2^+ + a_3
\sigma_1^+\sigma_2^+,
\end{gather*}
which acting on vacuum state $|{00}\rangle$ yields,
$a_0|00\rangle+a_1|10\rangle+a_2|01\rangle+a_3|11\rangle$. So one
can take the coef\/f\/icients  such that the state becomes MES. As a
simple example taking $a_0=a_3=\frac{1}{\sqrt{2}}$ and $a_1=a_2=0$
yields Bell state $|\Phi^+\rangle$.

\subsection{Entangled supper coherent state}
The study of entanglement in a system involve both bosons and
fermions remains one of the most challenging problems in quantum
information science \cite{Duff,Castellani}. Here we want to show
that in some cases it is possible to construct MES using
superposition of bosonic and fermionic coherent states.  Fermions
are described in the anti-commuting Grassmann coordinate space,
consequently one must use super-Hilbert space to study both bosons
and fermions. A bosonic coherent state can be def\/ined as eigenstate
of the annihilation operator
\begin{gather*}
b|\alpha\rangle=\alpha|\alpha\rangle,
\end{gather*}
where $\alpha$ is a complex number, and $b$ is annihilation operator
for the bosonic coherent state
\begin{gather*}
|\alpha\rangle=e^{\frac{-|\alpha|^2}{2}}
\sum_{n=0}^\infty\frac{\alpha^n}{\sqrt{n!}}|n\rangle=D(\alpha)|0\rangle,
\end{gather*}
and $D(\alpha)$, is displacement operator
\begin{gather*}
D(\alpha):=\exp(a^\dag\alpha-\alpha^*a).
\end{gather*}
We can also express that
\begin{gather*}
\langle\alpha|\beta\rangle=e^{-\frac{1}{2}(|\alpha|^2+|\beta|^2-2\alpha^*\beta)}.
\end{gather*}
With this background, let $|\alpha\rangle$ and $|\beta\rangle$ be two
orthogonal bosonic coherent states, which is possible in the limit
$\alpha\rightarrow\infty$ and $\beta\rightarrow0$. Therefore in the
space spanned by two coherent states one can take
$|\alpha\rangle\equiv|1\rangle_b$ and
$|\beta\rangle\equiv|0\rangle_b$. On the other hand in the space of
fermionic system we can describe the states $|1\rangle_f$ and
$|0\rangle_f$ integrating on fermionic coherent states. Now consider
the following super coherent states  belonging to the Hilbert space
$\mathcal{H}_{\rm fermion}\otimes \mathcal{H}_{\rm boson}$
\begin{gather*}
\lim_{\alpha\rightarrow\infty}\int
d\theta\frac{\theta}{\sqrt{2}}|\theta\rangle|\alpha\rangle\mp
\lim_{\beta\rightarrow0}\int
d\theta\frac{1}{\sqrt{2}}|\theta\rangle|\beta\rangle=\frac{1}{\sqrt{2}}(|0\rangle_f|1\rangle_b\pm|1\rangle_f|0\rangle_b),
\end{gather*}
and likewise
\begin{gather*}
\lim_{\alpha\rightarrow0}\int
d\theta(\frac{-1}{\sqrt{2}})|\theta\rangle|\alpha\rangle\pm
\lim_{\beta\rightarrow\infty}\int
d\theta\frac{\theta}{\sqrt{2}}|\theta\rangle|\beta\rangle=\frac{1}{\sqrt{2}}(|0\rangle_f|0\rangle_b\pm|1\rangle_f|1\rangle_b).
\end{gather*}
The right hand sides of the above equations may be interpreted as
two partite  MESs in the  Hilbert space
$\mathcal{H}_{\rm fermion}\otimes \mathcal{H}_{\rm boson}$. We not that, if
$\alpha\rightarrow\beta$, then we obtain  separable states, i.e.,
\begin{gather*}
\lim_{\alpha\rightarrow\beta}\int
d\theta\frac{\theta}{\sqrt{2}}|\theta\rangle|\alpha\rangle\mp \int
d\theta\frac{1}{\sqrt{2}}|\theta\rangle|\beta\rangle=\frac{1}{\sqrt{2}}(|0\rangle_f\pm|1\rangle_f)|\alpha\rangle.
\end{gather*}
Of course the other way to create the orthogonal basis in the
bosonic Hilbert space is
\begin{gather*}
|0\rangle_b=|\alpha\rangle, \qquad
|1\rangle_b=\frac{|\beta\rangle-\langle\alpha|\beta\rangle|\alpha\rangle}{N_1},
\qquad \text{where}\qquad N_1=\sqrt{1-|\langle\alpha|\beta\rangle|}.
\end{gather*}
Thus instead of taking limit  we can use the above bases to
construct MESs in $\mathcal{H}_{\rm fermion}\otimes
\mathcal{H}_{\rm boson}$,  however in spite of the fact that it belongs
to $\mathcal{H}_{\rm boson}$ $|1\rangle_b$ is not a coherent state
anymore.

The approach used here is somewhat dif\/ferent from the method used by
Castellani et~al.\ in~\cite{Castellani}.  They established the
superqubits states in composed Hilbert space of the form
$H_{\rm boson}\oplus H_{\rm fermion}$, with general form
\begin{gather*}
|\psi\rangle=\sum_{i}^n b_{i}|B_{i}\rangle+\theta\sum_{i}^n
f_{i}|F_{i}\rangle,
\end{gather*}
with $n$ bosonic states $|B_{i}\rangle$ and $n$ fermionic states
$|F_{i}\rangle$, combined via a single Grassmann coordinate $\theta$
and then normalized it with suitable weight function. For identical
particles they symmetrized tensor product superqubits to produce the
entangled states such as
\begin{gather*}
|\psi\rangle\otimes|\psi'\rangle=\sum_{i,j}^n
b_{i}b'_{j}|B_{i}B_{j}\rangle+\theta\sum_{i,j}^n
(b_{i}f'_{j}+b'_{i}f_{j})|B_{i}F_{j}\rangle.
\end{gather*}
Basically  there are  some dif\/ferences between  these constructions.
For example, we begin to establish fermionic state as an eigenstate
of the fermionic annihilation operator, while  the above superqubit
state  has no connection to such operator at all. On the other hand
there is no need to normalize  GCSs beforehand since the
normalization factor can be included in the weight function.

\section{Conclusion}\label{section4}

In conclusion, we have investigated the entanglement of
multi-partite Grassmannian coherent states (GCS)  described by
anti-commuting Grassmann numbers. This task is achieved  by
integration over Grassmann numbers with taking appropriate weight
functions. In other words we established a relation between GCS and
pure entangled states which are treated almost separately. The
construction of {\bf GHZ}, {\bf W}, Bell, cluster type and
bi-separable states and also generalization of entanglement for $n$
level GCSs was developed.
 It is shown that for three level systems, the
 creation and
 annihilation operators
$b$ and  $b^\dag$ together with $b_{z}:=[b,b^{\dag}]_{q}$ form a~closed deformed algebra $SU_{q}(2)$. The similar  closed algebra
was obtained for three operators~$b^2$,~$b^{{\dag}^2}$ and
$b'_{z}:=[b^2,b^{{\dag}^2}]$,  which  one may tempt to
 def\/ine  Grassmannian squeezed state and  create some MESs using these squeezed states.
For three level systems, combination of Grassmannian coherent and
squeezed states yields other MESs which was not possible to be
obtained using  GCSs with just one Grassmann number before.

Finally combining the Grassmann and bosonic coherent states we
 construct maximal entangled super coherent states.

\pdfbookmark[1]{References}{ref}
\LastPageEnding

\end{document}